\documentclass[a4paper,british,english]{IEEEtran}
\usepackage[T1]{fontenc}
\usepackage[latin9]{inputenc}
\usepackage{verbatim}
\usepackage{amsmath}
\usepackage{amsthm}
\usepackage{amssymb}
\usepackage{graphicx}

\makeatletter

\theoremstyle{plain}
\newtheorem{thm}{\protect\theoremname}
\theoremstyle{remark}
\newtheorem{rem}[thm]{\protect\remarkname}

\renewcommand{\fnum@figure}{Fig.~\thefigure}
\usepackage{cite}

\usepackage{amsthm}
\newtheorem{thm2}{Theorem}
\newtheorem{cor2}{Corollary}

\usepackage{layout}

\makeatother

\usepackage{babel}
\addto\captionsbritish{\renewcommand{\remarkname}{Remark}}
\addto\captionsbritish{\renewcommand{\theoremname}{Theorem}}
\addto\captionsenglish{\renewcommand{\remarkname}{Remark}}
\addto\captionsenglish{\renewcommand{\theoremname}{Theorem}}
\providecommand{\remarkname}{Remark}
\providecommand{\theoremname}{Theorem}

\begin{document}

\title{Novel Performance Analysis of Network Coded Communications in Single-Relay
Networks}

\author{Evgeny Tsimbalo, Andrea Tassi, Robert J. Piechocki\\
Department of Electrical and Electronic Engineering, University of
Bristol, Bristol, UK\\
Email: \{e.tsimbalo, a.tassi, r.j.piechocki\}@bristol.ac.uk}
\maketitle
\begin{abstract}
In this paper, we analyze the performance of a single-relay network
in which the reliability is provided by means of Random Linear Network
Coding (RLNC). We consider a scenario when both source and relay nodes
can encode packets. Unlike the traditional approach to relay networks,
we introduce a passive relay mode, in which the relay node simply
retransmits collected packets in case it cannot decode them. In contrast
with the previous studies, we derive a novel theoretical framework
for the performance characterization of the considered relay network.
We extend our analysis to a more general scenario, in which coding
coefficients are generated from non-binary fields. The theoretical
results are verified using simulation, for both binary and non-binary
fields. It is also shown that the passive relay mode significantly
improves the performance compared with the active-only case, offering
an up to two-fold gain in terms of the decoding probability. The proposed
framework can be used as a building block for the analysis of more
complex network topologies. 
\end{abstract}

\section{Introduction\label{sec:Introduction}}

Random Linear Network Coding (RLNC) has been originally proposed by
R. Ahlswede~\textit{et al.}~\cite{RR0} as a way to improve the
communication throughput over a wired multi\nobreakdash-hop network.
Nowadays, the RLNC principle is being applied in several application
domains~\cite{medard2012network}. This paper focuses on the adoption
of RLNC as a way to improve the reliability over a network~\cite{RR1,RR3}.
The principle underneath that RLNC application has been presented
in \cite{RR1} and can be summarized as follows. Instead of directly
transmitting a set of \emph{source packets}, the source node generates
and transmits a stream of \emph{coded packets} to one or more destination
nodes. Each coded packet is obtained as a linear combination of the
source packets. The destination nodes recover the set source packets
as soon as they collect a number of linearly independent coded packets
equal to the number of the source packets.

Generally, multi-hop relay networks interconnecting one or more source
nodes to several destination nodes have been extensively investigated~\cite{R0,R1,R2}.
In particular,~\cite{R0,R1} refer to system models where a source
node transmits a packet stream to multiple destination nodes, via
several intermediate relay nodes. In that tree\nobreakdash-based
topology, network flows are combined by the relay nodes to achieve
the min\nobreakdash-cut max\nobreakdash-flow~\cite{medard2012network}
between the source and each destination node. In fact,~\cite{R0,R1}
only refer to the classic binary RLNC principle as a way to efficiently
combine network information flows. Unlike~\cite{R0} and \cite{R1},~\cite{R2}
characterizes the performance of a multi-relay network where RLNC
operations are performed over a generic Galois Field (GF) composed
by an arbitrary number of elements. Also in that case, we observe
that~\cite{R2} refers to a tree\nobreakdash-based network topology
as in~\cite{R0,R1}. Differently from~\cite{R0,R1,R2}, our system
model adopts the RLNC principle as a way to improve communication
reliability. Hence, both the source and relay nodes are supposed to
transmit streams of coded packets towards the destination node.

The RLNC principle has also been adopted in more complex network topologies,
which may not necessarily resemble a direct acyclic graph, such as
a mobile ad-hoc network~\cite{XOR,R3}. In that case, intermediate
nodes are supposed to combine multiple incoming network flows and
typically relay one combined flow towards the next intermediate node
or the destination node. Also in this application domain, the goal
is that of achieving the min-cut max-flow and not that of improving
the reliability of communications. Another key difference between
the aforementioned papers and ours is that we assume that the source
node can communicate with the destination node directly, not only
via a relay node. In other words, we refer to a system model where
the destination node can be both a one-hop and two-hop neighbour of
the source node, which can be observed in some ``casual'' domestic
network deployments \cite{sphere}.

Similar to our RLNC application over a relay network is what has been
proposed in~\cite{Amjad2015}. In that paper, the source node transmits
streams of coded packets to both a relay node and destination nodes.
However, the relay node transmits a newly generated stream of coded
packets \emph{only if} it can recover what has been previously transmitted
by the source node. Unlike~\cite{Amjad2015}, if the relay node has
not been able to recover the original source message, we assume that
it simply relays the set of coded packets that it has been able to
collect. In the reminder of the paper, we will show how that improves
the overall communication reliability by up to two times. We also
observe that~\cite{Amjad2015} provides an approximated performance
model. Even though we refer to a simpler system model, we derive a
novel theoretical framework for a single-relay network, which is exact
for any GF, and explicitly adopts the RLNC principle as a way to improve
communication reliability. The result can be used as a building block
for the analysis of more complex networks.

The paper is organized as follows. Section~\ref{sec:System-model}
presents the considered system model. Section~\ref{sec:Sec3} describes
the proposed theoretical framework based on an exact performance characterization
of RLNC over a broadcast network topology. Numerical results are presented
in Section~\ref{sec:Sec4}, while our conclusions are drawn in Section~\ref{sec:Conclusions}.

\section{System Model and Background\label{sec:System-model}}

\begin{figure}
\begin{centering}
\includegraphics{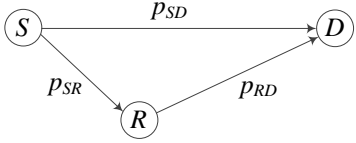}
\par\end{centering}

\caption{Block diagram of a single-user single-relay network with packet error
probabilities $p_{SD}$, $p_{SR}$ and $p_{RD}$.\label{fig:Single-user-single-relay-network}}

\end{figure}

A single-user single-relay network is depicted in Fig.~\ref{fig:Single-user-single-relay-network}.
The network consists of a source node $S$, relay $R$ and destination
$D$, and the goal is to transmit a message comprising of $K$ equal-size
packets from $S$ to $D$. To this end, the source node encodes the
$K$ message packets by using RLNC, such that each coded packet is
a linear combination of the original packets with the coefficients
drawn uniformly at random over a $GF(q)$, where $q$ is the field
size. In total, $S$ transmits $N_{S}\geq K$ coded packets. As proposed
in \cite{AT2}, we assume that all receiving nodes have a knowledge
of seeds used to generate the coded packets they receive, such that
the coding vector of each packet can be re-generated.

Let $p_{SD}$, $p_{SR}$ and $p_{RD}$ denote the Packet Error Probabilities
(PEPs) of the links connecting $S$ with $D$, $S$ with $R$, and
$R$ with $D$, respectively. As is traditional in relay networks,
the communication is performed in two stages. At the first stage,
$S$ transmits coded packets to $R$ and $D$. Both $R$ and $D$
receive a number of packets, for which they can restore the corresponding
coding vectors. The latter are stacked together horizontally to form
an $M_{R}\times K$ coding matrix $\mathbf{C}_{R}$ at the relay node
and an $M_{D}\times K$ coding matrix $\mathbf{C}_{D}$ at the destination
node. Since $R$ and $D$ may receive the same packets from $S$,
the matrices may have common rows. Let $M_{RD}$ denote the number
of such common rows. At this point, the destination may have enough
coding packets to make its coding matrix $\mathbf{C}_{D}$ full-rank,
thus being able to decode the original message without any assistance
from $R$.

At the second stage, it is checked if the relay coding matrix $\mathbf{C}_{R}$
is full rank. If so, $R$ decodes the original $K$ packets and re-encodes
them using newly generated random coefficients from $GF(q)$ into
$N_{R}$ packets and transmits them to $D$. In general, $N_{R}\neq N_{S}$.
We call this case the \textit{active relay} mode. If the relay node
cannot decode the source packets, it simply re-transmits the $M_{R}$
packets to the destination node, which corresponds to the \textit{passive
relay} mode. In either mode, we denote $M_{D}'$ as the number of
coding vectors reached $D$ from $R$, and $\mathbf{C}_{D}^{\prime}$
as the updated $M_{D}+M_{D}^{\prime}\times K$ coding matrix at~$D$. 

The described relay network is different from the one proposed previously
in \cite{Amjad2015}, since in the latter the relay node transmits
only if it can decode packets from the source node. Naturally, the
passive relay mode should improve the decoding probability at $D$
in cases when $R$ fails to decode. In addition, the described system
and its analysis can be straightforwardly extended to a network with
multiple sources, in which each source has an independent communication
channel. This again contrasts with \cite{Amjad2015}, where the relay
node uses packets from both sources at the encoding stage, thus introducing
correlation between the sources.

\subsection{Theoretical Background}

We now provide some background results on RLNC which we will use in
our analysis. 

The number of full-rank $m\times k$ matrices generated over $GF(2)$,
with $m\geqslant k$, is given by \cite{Ferreira2013}
\begin{equation}
F(m,k)=\prod_{i=0}^{k-1}(2^{m}-2^{i})=2^{mk}\prod_{i=0}^{k-1}(1-2^{i-m}).\label{eq:NumberOfFullRank}
\end{equation}
Based on that, the number of matrices of the same size that have rank
$r\leq k$ can be calculated as
\begin{equation}
G(m,k,r)=\frac{F(m,r)F(k,r)}{F(r,r)}.\label{eq:NumOflRankr}
\end{equation}
The probability of a random $m\times k$ matrix having full rank can
be obtained by dividing (\ref{eq:NumberOfFullRank}) by the number
of all possible $m\times k$ matrices:\vspace*{-4mm}

\begin{equation}
\mathbb{P}(m,k)\triangleq2^{-mk}F(m,k).\label{eq:ProbFullRank}
\end{equation}
Similarly, the probability of a random $m\times k$ matrix having
rank $r\leq k$ is given by
\begin{equation}
\mathbb{P}_{r}(m,k)\triangleq2^{-mk}G(m,k,r).\label{eq:ProblRankr}
\end{equation}
In a general case, when the elements are generated from $GF(q)$,
$q\geq2$, (\ref{eq:ProbFullRank}) can be rewritten by simply replacing
$2$ with $q$ (see, for example, \cite{RR3}):
\begin{equation}
\mathbb{P}(m,k)=\prod_{i=0}^{k-1}(1-q^{i-m}).\label{eq:ProbFullRank_q}
\end{equation}
Furthermore, following the same train of thought used to obtain (\ref{eq:NumOflRankr})
in the binary case, the probability (\ref{eq:ProblRankr}) can be
\foreignlanguage{british}{generalized} to the non-binary case as follows:\setlength{\arraycolsep}{0.14em}
\begin{eqnarray}
\mathbb{P}_{r}(m,k) & = & q^{-mk}G(m,k,r)\label{eq:ProbRankr_q}\\
 & = & \frac{1}{q^{(m-r)(k-r)}}\prod_{i=0}^{r-1}\frac{(1-q^{i-m})(1-q^{i-k})}{1-q^{i-r}}.\nonumber 
\end{eqnarray}
\setlength{\arraycolsep}{5pt}%
{} 

Consider now the application of RLNC to a point-to-point link, with
a source node encoding $K$ source packets and transmitting $N$ coded
packets to the destination. The probability of successful decoding
for such link characterized by the PEP $p$ can be given by \cite{Amjad2015}
\begin{equation}
P_{ptp}(N,K,p)=\sum_{M=K}^{N}B(M,N,p)\mathbb{P}(M,K),\label{eq:Pptp}
\end{equation}
where $B(M,N,p)$ is the probability mass function (PMF) of the binomial
distribution:
\begin{equation}
B(M,N,p)=\binom{N}{M}(1-p)^{M}p^{N-M}.\label{eq:PMF_binom}
\end{equation}

In addition to the binomial distribution, we will also need its generalized
version - the multinomial distribution \cite{Papoulis84}. The PMF
of such distribution describes the probability of a particular combination
of numbers of occurrences of $k$ possible mutually exclusive outcomes
out of $n$ independent trials and is given by\setlength{\arraycolsep}{0.14em}\vspace*{-2mm}
\begin{equation}
f(\mathbf{x};n,\mathbf{\theta})=\alpha(\mathbf{x})\theta_{1}^{x_{1}}\cdots\theta_{k}^{x_{k}},\quad\mathrm{for}\,\sum_{i=1}^{k}x_{i}=n,\label{eq:PMF_multinom}
\end{equation}
\setlength{\arraycolsep}{5pt}where $\mathbf{x}=(x_{1},\ldots,x_{k})$
is a combination of numbers of occurrences, such that the sum of them
equals $n$, ${\mathbf{\theta}=(\theta_{1},\ldots,\theta_{k})}$ are
probabilities of each outcome and\setlength{\arraycolsep}{0.14em}
\begin{eqnarray}
\alpha(\mathbf{x}) & = & \frac{n!}{x_{1}!x_{2}!\ldots x_{k}!}\nonumber \\
 & = & \binom{n}{x_{1}}\binom{n-x_{1}}{x_{2}}\ldots\binom{n-\sum_{i=1}^{k-2}x_{i}}{x_{k-1}}\label{eq:MultinomCoeff}
\end{eqnarray}
\setlength{\arraycolsep}{5pt}is the multinomial coefficient.

\section{Proposed Theoretical Framework\label{sec:Sec3}}

Our goal is to derive the probability of successful decoding for the
single-relay network described in the previous section. In contrast
with \cite{Amjad2015}, we aim to derive the exact formulation, by
taking into account the fact that the coding matrices of $R$ and
$D$ are correlated due to the presence of common rows. But first,
we establish some preliminary results.

\subsection{Preliminaries\label{sub:Preliminaries}}

Consider a block matrix $\mathbf{X}$ composed of two vertically concatenated
random matrices $\mathbf{A}$ and $\mathbf{B}$ with dimensions $a\times k$
and $b\times k$ generated over $GF(q)$:\vspace*{-2mm}
\begin{equation}
\mathbf{X}=\left(\begin{array}{c}
\mathbf{A}\\
\mathbf{B}
\end{array}\right).\label{eq:MatX}
\end{equation}
\vspace*{-2mm}

\begin{cor2} The probability of matrix $\mathbf{X}$ being full-rank is given by\footnote{It should be noted that a similar result was implicitly obtained in \cite{Amjad2015}, albeit with the summation index starting from zero. Here, we highlight the result and obtain a more accurate starting value for the summation index, thus eliminating unnecessary zero terms under the summation.} 
\vspace*{-2mm}
\begin{equation} \Pr[\mathrm{rank}\,\mathbf{X}=k]=\sum_{i=\max(0,k-b)}^{\min(a,k)}\mathbb{P}_{i}(a,k)\mathbb{P}(b,k-i).\label{eq:ProbXFullRank} \end{equation} \end{cor2}

\begin{IEEEproof}
This corollary is based on Theorem~2 in \cite{Ferreira2013}, where
the authors proved the result for a more general case, when $\mathbf{X}$
is a block angular matrix. Here, we follow the same train of thought.
Suppose that $\mathbf{A}$ has rank $i$ which is upper-bounded by
its smallest dimension, and hence has $i$ linearly independent columns.
$\mathbf{X}$ is then full-rank if $\mathbf{B}$ has $k-i$ linearly
independent columns, with the rest of its columns selected in an arbitrary
way. The number of ways $\mathbf{A}$ can be chosen to have rank $i$
is $G(a,k,i)$. The $k-i$ linearly independent columns of $\mathbf{B}$
can be selected in $F(b,k-i)$ ways, while the $i$ arbitrary columns
of $\mathbf{B}$ can be selected in $q^{ib}$ ways. Finally, the total
number of possible matrices $\mathbf{X}$ is $q^{(a+b)k}$. This yields
the following:\setlength{\arraycolsep}{0.0em}
\begin{eqnarray}
\Pr[\mathrm{rank}\,\mathbf{X} =k]&{}={}&q^{-(a+b)k}\sum_{i}G(a,k,i)F(b,k-i)q^{ib}\nonumber \\  
&{}={}&\sum_{i}q^{-ak}G(a,k,i)q^{-b(k-i)}F(b,k-i)\nonumber \\  
&{}={}&\sum_{i}\mathbb{P}_{i}(a,k)\mathbb{P}(b,k-i).\label{eq:PrRank}
\end{eqnarray}
\setlength{\arraycolsep}{5pt}The starting value of $i$ in the summation is based on the fact that
the number of rows $b$ in matrix $\mathbf{B}$ should be larger than
$k-i$ to provide non-zero $\mathbb{P}(b,k-i)$; if $b>k$, then $i$
should start with zero.
\end{IEEEproof}
The corollary will be utilized to derive a probability of two matrices
with common rows being simultaneously full-rank. We introduce the
following theorem:

\begin{thm2} The probability of two random matrices $\mathbf{X}_{1}$ and $\mathbf{X}_{2}$ generated over $GF(q)$ with dimensions $m_{1}\times k$ and $m_{2}\times k$, $m_{1},m_{2}\geq k$, and $m_{12}$ common rows being simultaneously full rank is given by\setlength{\arraycolsep}{0.0em} \begin{eqnarray} \mathbb{P}^{*}(\mathbf{m},k)&{}={}&\sum_{i}\mathbb{P}_{i}(m_{12},k)\mathbb{P}(m_{1}-m_{12},k-i)\notag\\ &&{}\cdot\mathbb{P}(m_{2}-m_{12},k-i),\label{eq:ProbFullRankBoth}  \end{eqnarray} \setlength{\arraycolsep}{5pt}where $\mathbf{m}=(m_{1},m_{2},m_{12})$ and the summation is performed over the values of $i$ from $\max(0,k-m_{1}+m_{12},k-m_{2}+m_{12})$ to $\min(m_{12},k)$.\end{thm2}

\begin{IEEEproof}
The proof follows directly from (\ref{eq:ProbXFullRank}) by noting
that $m_{1}-m_{12}$ rows of $\mathbf{X}_{1}$ are statistically independent
from $m_{2}-m_{12}$ rows of $\mathbf{X}_{2}$. The starting value
of $i$ is chosen to exclude unnecessary summation terms when one
of the last two probabilities under the summation is zero.
\end{IEEEproof}
Let us now consider a three-node network in which one of the nodes
multicasts $N$ randomly encoded packets derived from $K$ source
packets to the other two nodes, with the PEPs $p_{1}$ and $p_{2}$
grouped into $\mathbf{p}=(p_{1},p_{2})$. Let $M_{1}$ and $M_{2}$
denote possible numbers of coding vectors received by the destinations,
and $M_{12}$ denote a possible number of common coding vectors among
them. To simplify the notation, we define $\mathbf{M}=(M_{1},M_{2},M_{12})$.
We establish the following:

\begin{thm2} \label{thm:Multicast}The probability of successful decoding for a two-destination multicast network defined by parameters $N,K$ and $\mathbf{p}$ is given by \begin{equation} P_{M}(N,K,\mathbf{p})=\sum_{\mathbf{M}}B^{*}(\mathbf{M},N,\mathbf{p})\mathbb{P}^{*}(\mathbf{M},K),\label{eq:ProbMulticast} \end{equation} where\setlength{\arraycolsep}{0.14em} \begin{eqnarray} B^{*}(\mathbf{M},N,\mathbf{p}) & = & \binom{N}{M_{12}}\binom{N-M_{12}}{M_{1}-M_{12}}\binom{N-M_{1}}{M_{2}-M_{12}}\nonumber \\  &  & \cdot(1-p_{1})^{M_{1}}p_{1}^{N-M_{1}}\nonumber \\  &  & \cdot(1-p_{2})^{M_{2}}p_{2}^{N-M_{2}}\label{eq:Bstar} \end{eqnarray} \setlength{\arraycolsep}{5pt}and the summation is performed over the following values: \begin{equation} \begin{cases} M_{1},M_{2}=K,\ldots,N;\\ M_{12}=\max(0,M_{1}+M_{2}-N),\ldots,\min(M_{1},M_{2}). \end{cases}\hspace{-2mm}\label{eq:Limits} \end{equation} \end{thm2}

\begin{IEEEproof}
Let the received coding vectors be arranged into matrices $\mathbf{C}_{1}$
and $\mathbf{C}_{2}$ with dimensions $M_{1}\times K$ and $M_{2}\times K$.
Similarly to the point-to-point communication as in (\ref{eq:Pptp}),
the probability of successful decoding at both destinations can be
marginalized over all possible values of $\mathbf{M}$ as follows:\begin{equation}
P_{M}(N,K,\mathbf{p})=\sum_{\mathbf{M}}\Pr[\mathbf{M}]\Pr[\mathrm{rank}\,\mathbf{C}_{1},\mathbf{C}_{2}=K|\mathbf{M}].\label{eq:ProbMulticast2}
\end{equation}It can be immediately observed that the second term under the summation
in (\ref{eq:ProbMulticast2}) can be expressed using (\ref{eq:ProbFullRankBoth})
as $\mathbb{P}^{*}(\mathbf{M},K)$. To calculate the first term, we
can model the transmission of $N$ packets over two lossy channels
as $N$ independent trials each having one of the four mutually exclusive
outcomes: the packet is received by both destinations, by one of them
or by neither of them. The probability in question is that of a particular
combination of numbers of occurrences of these events, which is described
by the PMF of the multinomial distribution, as in (\ref{eq:PMF_multinom}),
with $\mathbf{x}=(M_{12},M_{1}-M_{12},M_{2}-M_{12},N-M_{1}-M_{2}+M_{12})$
and probabilities $(1-p_{1})(1-p_{2})$, $(1-p_{1})p_{2}$, $p_{1}(1-p_{2})$
and $p_{1}p_{2}$, respectively. By substituting these into (\ref{eq:PMF_multinom})
and grouping, it can be shown that $\Pr[\mathbf{M}]=B^{*}(\mathbf{M},N,\mathbf{p})$. 

As regards the values of $\mathbf{M}$ over which in the summation
in (\ref{eq:ProbMulticast2}) is performed, they should be selected
such that both destinations receive at least $K$ coding vectors.
The starting value of $M_{12}$ can be defined as follows:
\begin{equation}
M_{12}\geq\begin{cases}
0, & N\geqslant M_{1}+M_{2}\\
M_{1}+M_{2}-N, & N<M_{1}+M_{2}
\end{cases}.\label{eq:}
\end{equation}
 At the same time, $M_{12}$ cannot exceed either $M_{1}$ or $M_{2}$.
The limits (\ref{eq:Limits}) follow, which concludes the proof.\end{IEEEproof}
\begin{rem}
One may observe the resemblance between the results for the point-to-point
and multicast scenarios ((\ref{eq:Pptp}) and (\ref{eq:ProbMulticast}),
respectively), implying that the latter is a generalization of the
former when the number of receivers is larger than one. 
\end{rem}
\vspace{0mm}

\begin{rem}
\label{rem:Approximation}It should also be noted that (\ref{eq:ProbMulticast})
can be approximated as a product of two point-to-point probabilities
(\ref{eq:Pptp}) by letting $M_{12}=0$:
\begin{equation}
P_{M}(N,K,\mathbf{p})\cong P_{ptp}(N,K,p_{1})P_{ptp}(N,K,p_{2}).\label{eq:Pmulticast_approx}
\end{equation}
 The higher a potential number of common rows, the more loose the
approximation becomes. In the ultimate case, when one of the coding
matrices consists of the common rows only, the decoding probability
of the multicast network converges to that of the point-to-point link.
Therefore, (\ref{eq:Pmulticast_approx}) can be used as a lower bound,
which was employed in \cite{Amjad2015} to characterize the performance
of a relay network. In contrast, we show in the next section how the
proposed analysis of the multicast network can be used to derive the
exact decoding probability of the relay network. 
\end{rem}

\subsection{Relay Network Modelling\label{sub:Relay-network}}

We now turn our attention towards the single-relay network described
in Section~\ref{sec:System-model}. The successful decoding event
at the destination can be decomposed into three independent sub-events
depending on a type of communication between $S$ and~$D$:

\subsubsection{Direct communication}

In this mode, $D$ can decode the source packets after the first stage.
The corresponding probability of successful decoding is that of a
point-to-point link (\ref{eq:Pptp}), with $N_{S}$ transmitted packets
and PEP $p_{SD}$:
\begin{equation}
P_{R,1}=P_{ptp}(N_{S},K,p_{SD}).\label{eq:Prelay1}
\end{equation}

\subsubsection{Relay-assisted communication, active relay}

In this mode, $D$ cannot decode the message after the first stage,
while $R$ can. During the second stage of transmission, the relay
node re-encodes the source message and transmits $N_{R}$ coded packets
to the destination node. %
Since these packets are generated at random, the number of common
coding vectors $M_{RD}$ between $R$ and $D$ remains the same. Employing
the same approach used to derive the decoding probability of the two-destination
multicast network (\ref{eq:ProbMulticast}), the decoding probability
for the relay network in the active relay mode can be written as\setlength{\arraycolsep}{0.14em}
\begin{eqnarray}
P_{R,2} & = & \sum_{\mathbf{M}}B^{*}(\mathbf{M},N_{S},\mathbf{p})\label{eq:Prelay2-0}\\
 &  & \cdot\Pr[\mathrm{rank}\,\mathbf{C}_{R},\mathbf{C}_{D}^{\prime}=K,\mathrm{rank}\,\mathbf{C}_{D}<K|\mathbf{M}],\nonumber 
\end{eqnarray}
\setlength{\arraycolsep}{5pt}where $\mathbf{M}=(M_{R},M_{D},M_{RD})$,
$\mathbf{p}=(p_{SR},p_{SD})$. Using the results of Theorem~\ref{thm:Multicast}
and noting that $D$ may not receive any packets from $S$ at all,
the values over which the summation is performed can be summarized
as follows:
\begin{equation}
\begin{cases}
M_{R}=K,\ldots,N_{S};\,M_{D}=0,\ldots,N_{S};\\
M_{RD}=\max(0,M_{R}+M_{D}-N_{S}),\ldots,\min(M_{R},M_{D}).
\end{cases}\label{eq:LimitsR2}
\end{equation}
 The probability term under the summation in (\ref{eq:Prelay2-0})
depends on the number of received packets from the relay node $M_{D}^{\prime}$.
For a given $M_{D}^{\prime}$, this probability can be derived as
follows:\setlength{\arraycolsep}{0.0em}
\begin{eqnarray}
&&\Pr[\mathrm{rank}\,\mathbf{C}_{R},\mathbf{C}_{D}^{\prime} = K,\mathrm{rank}\,\mathbf{C}_{D}<K|\mathbf{M},M_{D}^{\prime}]\nonumber \\
&{}={}& \Pr[\mathrm{rank}\,\mathbf{C}_{R},\mathbf{C}_{D}^{\prime}=K|\mathbf{M},M_{D}^{\prime}]\notag\\
&&{}- \Pr[\mathrm{rank}\,\mathbf{C}_{R},\mathbf{C}_{D} = K|\mathbf{M}]\notag\\
&{}={}& \mathbb{P}^{*}(\mathbf{M}^{\prime},K)-\mathbb{P}^{*}(\mathbf{M},K),\label{eq:Prelay2}
\end{eqnarray}
\setlength{\arraycolsep}{5pt}where $\mathbf{M}^{\prime}=(M_{R},M_{D}+M_{D}^{\prime},M_{RD})$ and
$\mathbb{P}^{*}$ is the probability of two matrices with common rows
being full-rank, as defined in (\ref{eq:ProbFullRankBoth}). Marginalizing
(\ref{eq:Prelay2}) over $M_{D}^{\prime}$ and substituting the result
into (\ref{eq:Prelay2-0}), the decoding probability in the active
relay mode can be expressed as follows:\setlength{\arraycolsep}{0.14em}\vspace*{-2mm}
\begin{eqnarray}
P_{R,2} & = & \sum_{\mathbf{M}}B^{*}(\mathbf{M},N_{S},\mathbf{p})\sum_{M_{D}^{\prime}=1}^{N_{R}}B(M_{D}^{\prime},N_{R},p_{RD})\nonumber \\
 &  & \cdot\left[\mathbb{P}^{*}(\mathbf{M}^{\prime},K)-\mathbb{P}^{*}(\mathbf{M},K)\right].\label{eq:Prelay2-1}
\end{eqnarray}
\setlength{\arraycolsep}{5pt}\vspace*{-2mm}

\subsubsection{Relay-assisted communication, passive relay}

In this mode, $R$ is not able to decode and just re-transmits the
packets it collects from $S$ to $D$. By analogy to (\ref{eq:Prelay2-0}),
the decoding probability in the passive relay mode can be written
as\setlength{\arraycolsep}{0.14em}
\begin{eqnarray}
P_{R,3} & = & \sum_{\mathbf{M}}B^{*}(\mathbf{M},N_{S},\mathbf{p})\nonumber \\
 &  & \cdot\Pr[\mathrm{rank}\,\mathbf{C}_{D}^{\prime}=K,\mathrm{rank}\,\mathbf{C}_{R},\mathbf{C}_{D}<K|\mathbf{M}],\label{eq:Prelay3_0}
\end{eqnarray}
\setlength{\arraycolsep}{5pt}where, using the results of Theorem~\ref{thm:Multicast}
and noting that $D$ may not receive any packets from $S$, while
$R$ should receive at least one packet from $S$, the values over
which the summation is performed are given by
\begin{equation}
\begin{cases}
M_{R}=1,\ldots,N_{S};\,M_{D}=0,\ldots,N_{S};\\
M_{RD}=\max(0,M_{R}+M_{D}-N_{S}),\ldots,\min(M_{R},M_{D}).
\end{cases}\label{eq:LimitsRelay3}
\end{equation}
Given the number of received packets from the relay node $M_{D}^{\prime}$,
the probability term in (\ref{eq:Prelay3_0}) can be derived as follows:\setlength{\arraycolsep}{0.14em}\vspace*{-4mm}

\begin{eqnarray}
 &  & \Pr[\mathrm{rank}\,\mathbf{C}_{D}^{\prime}=K,\mathrm{rank}\,\mathbf{C}_{R},\mathbf{C}_{D}<K|\mathbf{M},M_{D}^{\prime}]\nonumber \\
 & = & \Pr[\mathrm{rank}\,\mathbf{C}_{D}^{\prime}=K|\mathbf{M},M_{D}^{\prime}]-\Pr[\mathrm{rank}\,\mathbf{C}_{D}=K|\mathbf{M}]\nonumber \\
 &  & -\Pr[\mathrm{rank}\,\mathbf{C}_{R},\mathbf{C}_{D}^{\prime}=K,\mathrm{rank}\,\mathbf{C}_{D}<K|\mathbf{M},M_{D}^{\prime}]\nonumber \\
 & = & \mathbb{P}(M_{D}+M_{D}^{\prime},K)-\mathbb{P}^{*}(\mathbf{M}^{\prime\prime},K)\nonumber \\
 &  & +\mathbb{P}^{*}(\mathbf{M},K)-\mathbb{P}(M_{D},K),\label{eq:ProbRelay3}
\end{eqnarray}
\setlength{\arraycolsep}{5pt}where $\mathbf{M}^{\prime\prime}=(M_{R},M_{D}+M_{D}^{\prime},M_{RD}+M_{D}^{\prime})$.
Out of $M_{R}$ packets received by $R$, only $M_{R}-M_{RD}$ ones
unique to $R$ have value for $D$; the rest of them have already
been received from $S$. By marginalizing over $M_{D}^{\prime}$ appropriately
and denoting $M_{R}^{\prime}=M_{R}-M_{RD}$, the decoding probability
for the passive relay mode can be calculated as follows:\setlength{\arraycolsep}{0.14em}\vspace*{-2mm}
\begin{eqnarray}
P_{R,3} & = & \sum_{\mathbf{M}}B^{*}(\mathbf{M},N_{S},\mathbf{p})\sum_{M_{D}^{\prime}=1}^{M_{R}^{\prime}}B(M_{D}^{\prime},M_{R}^{\prime},p_{RD})\nonumber \\
 &  & \cdot[\mathbb{P}(M_{D}+M_{D}^{\prime},K)-\mathbb{P}(M_{D},K)\nonumber \\
 &  & -\mathbb{P}^{*}(\mathbf{M}^{\prime\prime},K)+\mathbb{P}^{*}(\mathbf{M},K)].\label{eq:ProbRelay31}
\end{eqnarray}
\setlength{\arraycolsep}{5pt}\vspace*{-4mm}

The overall decoding probability of the single-relay network can now
be obtained:
\begin{equation}
P_{R}=P_{R,1}+P_{R,2}+P_{R,3}.\label{eq:ProbRelayOverall}
\end{equation}

\section{Numerical Results\label{sec:Sec4}}

In this section, we investigate the performance of the described relay
network via simulation and compare the results with the theoretical
prediction. Simulation results are obtained using the Monte Carlo
method, with each point being the result of an average over $10^{5}$
iterations. 

\begin{figure}
\includegraphics[scale=0.6]{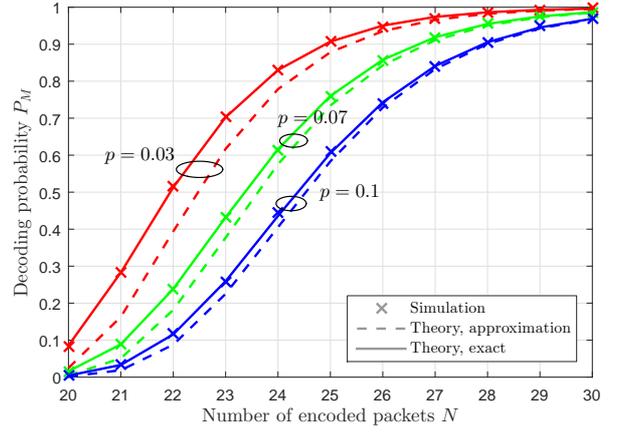}

\caption{Performance of the two-destination multicast network as a function
$N$, for a fixed $K=20$ and different values of $p$.\label{fig:Fig1}}

\end{figure}

We start by illustrating the importance of the exact formulation for
the decoding probability. To this end, we consider the performance
of the two-destination multicast network described in Section~\ref{sub:Preliminaries},
the theoretical characterization of which, as was shown earlier, plays
an important role in the analysis of the relay network. For simplicity,
both links are assumed to have the same PEP $p$. We use the exact
and approximated expressions for the decoding probability $P_{M}$
((\ref{eq:ProbMulticast}) and (\ref{eq:Pmulticast_approx}), accordingly)
and compare the resulting values with simulated ones. Fig.~\ref{fig:Fig1}
shows the results as a function of the number of transmitted packets
$N$, for different values of $p$ and a fixed $K=20$. While it can
be observed that the system performance is described accurately by
the exact formulation, the approximation has some inaccuracy, which
increases as $p$ becomes lower. The reason is that the number of
common packets received by both destination nodes grows when the links
reliability improves. The approximation gap is especially profound
for small values of $N$, which again can be explained by a higher
probability of receiving common packets. Finally, it can be seen that
the approximation is indeed a lower bound, as was predicted in Section~\ref{sub:Preliminaries}.

\begin{figure}
\includegraphics[scale=0.6]{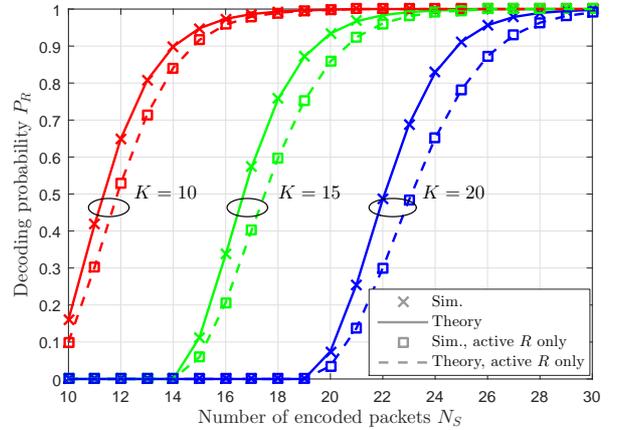}

\caption{Performance of the single-relay network as a function of $N_{S}$
for various values of $K$; $p_{SD}=0.3,p_{SR}=0.1$ and $p_{RD}=0.2$.\label{fig:Fig2}}
\end{figure}

We now consider the relay network described in Section~\ref{sub:Relay-network}.
For simplicity, hereafter we assume $N_{R}=N_{S}$. Fig.~\ref{fig:Fig2}
demonstrates simulated and theoretical values of the decoding probability
$P_{R}$ as a function of $N_{S}$, for various values of $K$. The
PEP values were selected to describe a typical relay network as follows:
$p_{SD}=0.3,p_{SR}=0.1,p_{RD}=0.2$. Fig.~\ref{fig:Fig2} also shows
the simulated and theoretical performance for a scenario when the
relay node is used in the active mode only, as in \cite{Amjad2015}.
It can be seen that the theoretical framework perfectly matches the
simulated results. At the same time, the proposed passive relay mode
significantly improves the performance, increasing the decoding probability
by up to $0.2$. The mode is especially beneficial for high values
of $K$, which is explained by a lower influence of the first two
terms in (\ref{eq:ProbRelayOverall}) when $K$ is increased. 

\begin{figure}
\includegraphics[scale=0.6]{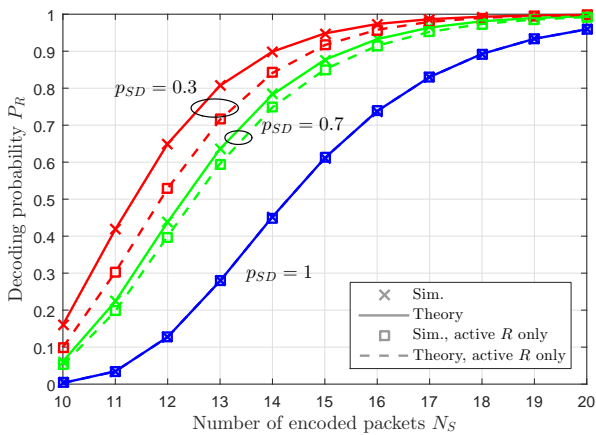}

\caption{Performance of the relay network as a function of $N_{S}$ for various
values of $p_{SD}$ and fixed $K=10,$ $p_{SR}=0.1$ and $p_{RD}=0.2$.\label{fig:Fig3}}

\end{figure}

Fig.~\ref{fig:Fig3} illustrates the performance of the relay network
as a function of $N_{S}$, but this time for different values of $p_{SD}$
and fixed $K=10$, $p_{SR}=0.1$ and $p_{SR}=0.2$. It is clear in
this scenario too that the theoretical results match the simulated
ones. At the same time, it can be observed that the effect of the
passive relay mode diminishes as the link between the source and destination
nodes becomes less reliable. Ultimately, when there is no direct connection
between $S$ and $D$ ($p_{SD}=1$), the latter can decode only if
$R$ can, hence the passive relay mode becomes redundant.

\begin{figure}
\includegraphics[scale=0.6]{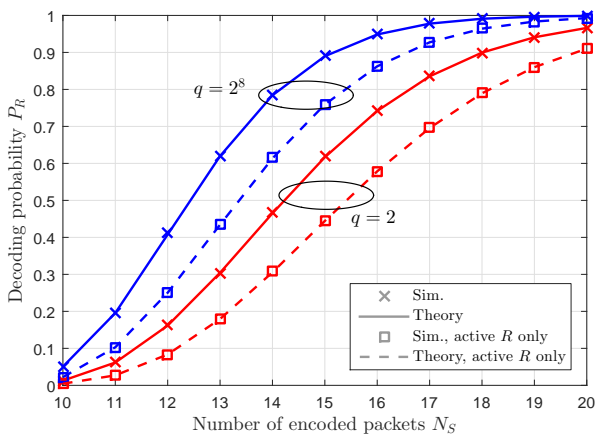}

\caption{Performance comparison between codes generated from binary ($q=2$)
and non-binary ($q=2^{8}$) codes; $K=10,$ $p_{SD}=0.5$, $p_{SR}=0.3$
and $p_{RD}=0.4$.\label{fig:Fig4}}

\end{figure}

Finally, the performance of the relay network is investigated for
a non-binary code in Fig.~\ref{fig:Fig4}. Here, we compare codes
generated from GF of size $q=2$ and $2^{8}$, for the same value
of $K=10$ and $p_{SD}=0.5$, $p_{SR}=0.3$, $p_{RD}=0.4$. It should
be noted that such a set of PEP values may describe a random deployment
scenario in which none of the links provides a satisfactory performance
level. It can be seen that the theoretical framework is valid for
the non-binary case as well. As expected, the non-binary code provides
a superior performance, offering the same decoding probability as
the binary code but requiring fewer encoded packets. The passive relay
mode is clearly beneficial for the non-binary code too, improving
the decoding probability by up to $0.2$.

\section{Conclusions\label{sec:Conclusions}}

In this work, we investigated the performance of a single-user single-relay
network in the context of RLNC, in a scenario where both source and
relay nodes encode packets. In contrast with the previous studies
related to relay networks, we derived the exact expression for the
decoding probability. In the process, we established some fundamental
results, such as the probability of two correlated matrices generated
from $GF(q)$ being full-rank and the decoding probability of a two-destination
multicast network. In addition, we proposed a passive relay mode,
in which the relay node re-transmits collected packets if it is not
able to decode them. Simulations showed that the established theoretical
framework accurately describes the performance of the relay network
not only for the binary, but for a non-binary code too. It was also
demonstrated that the proposed passive relay mode offers an additional
gain in the decoding probability of up to two times, compared with
the active-only scenario considered in the previous study. Future
work will deal with the generalization of the derived framework to
a multiple-relay network.

\section{Acknowledgments}

This work was performed under the SPHERE IRC funded by the UK Engineering
and Physical Sciences Research Council (EPSRC), Grant EP/K031910/1.

\bibliographystyle{ieeetr}
\bibliography{bib}

\begin{thebibliography}{10}

\bibitem{RR0}
R.~Ahlswede, N.~Cai, S.~Y.~R. Li, and R.~W. Yeung, ``{Network Information
  Flow},'' {\em {IEEE} Trans. Inf. Theory}, vol.~46, pp.~1204--1216, July 2000.

\bibitem{medard2012network}
M.~M\'edard and A.~Sprintson, {\em {Network Coding: Fundamentals and
  Applications}}.
\newblock Academic Press, Elsevier, 2012.

\bibitem{RR1}
M.~Ghaderi, D.~Towsley, and J.~Kurose, ``{Reliability Gain of Network Coding in
  Lossy Wireless Networks},'' in {\em Proc. IEEE INFOCOM 2008}, (Phoenix,
  Arizona, US-AZ), pp.~196--200, Apr. 2008.

\bibitem{RR3}
A.~Tassi, I.~Chatzigeorgiou, and D.~Vukobratovi\'c, ``{Resource-Allocation
  Frameworks for Network-Coded Layered Multimedia Multicast Services},'' {\em
  {IEEE} J. Sel. Areas Commun.}, vol.~33, pp.~141--155, Feb. 2015.

\bibitem{R0}
X.~Li, T.~Jiang, Q.~Zhang, and L.~Wang, ``{Binary Linear Multicast Network
  Coding on Acyclic Networks: Principles and Applications in Wireless
  Communication Networks},'' {\em {IEEE} J. Sel. Areas Commun.}, vol.~27,
  pp.~738--748, June 2009.

\bibitem{R1}
R.~Gummadi and R.~S. Sreenivas, ``{Relaying a Fountain Code Across Multiple
  Nodes},'' in {\em Proc. of IEEE ITW 2008}, (Porto, Portugal, PT),
  pp.~149--153, May 2008.

\bibitem{R2}
C.~F. Chiasserini, E.~Viterbo, and C.~Casetti, ``{Decoding Probability in
  Random Linear Network Coding with Packet Losses},'' {\em {IEEE} Commun.
  Lett.}, vol.~17, pp.~1--4, Nov. 2013.

\bibitem{XOR}
S.~Katti, H.~Rahul, W.~Hu, D.~Katabi, M.~M\'edard, and J.~Crowcroft, ``{XORs in
  the Air: Practical Wireless Network Coding},'' {\em {IEEE/ACM} Trans. Netw.},
  vol.~16, pp.~497--510, June 2008.

\bibitem{R3}
Y.~Qin, F.~Yang, X.~Tian, X.~Wang, H.~Luo, H.~Wang, and M.~Guizani, ``{Optimal
  Configuration of Network Coding in Ad Hoc Networks},'' {\em {IEEE} Trans.
  Veh. Technol.}, vol.~64, pp.~2001--2014, May 2015.

\bibitem{sphere}
X.~Fafoutis, E.~Tsimbalo, E.~Mellios, G.~Hilton, R.~Piechocki, and I.~Craddock,
  ``{A Residential Maintenance-Free Long-Term Activity Monitoring System for
  Healthcare Applications},'' {\em EURASIP Journal on Wireless Communications
  and Networking}, vol.~2016, no.~1, pp.~1--20, 2016.

\bibitem{Amjad2015}
A.~S. Khan and I.~Chatzigeorgiou, ``{Performance Analysis of Random Linear
  Network Coding in Two-Source Single-Relay Networks},'' in {\em Proc. of IEEE
  ICC 2015}, (London, United Kingdom, UK), pp.~991--996, June 2015.

\bibitem{AT2}
A.~Tassi, C.~Khirallah, D.~Vukobratovi\'c, F.~Chiti, J.~S. Thompson, and
  R.~Fantacci, ``{Resource Allocation Strategies for Network-Coded Video
  Broadcasting Services Over LTE-Advanced},'' {\em IEEE Trans. Veh. Technol.},
  vol.~64, pp.~2186--2192, May 2015.

\bibitem{Ferreira2013}
P.~J. S.~G. Ferreira, B.~Jesus, J.~Vieira, and A.~J. Pinho, ``{The Rank of
  Random Binary Matrices and Distributed Storage Applications},'' {\em {IEEE}
  Commun. Lett.}, vol.~17, no.~1, pp.~151--154, 2013.

\bibitem{Papoulis84}
A.~Papoulis, {\em Probability, Random Variables, and Stochastic Processes}.
\newblock New York: McGraw-Hill, 2nd~ed., 1984.

\end{thebibliography}

\end{document}